\documentclass[mathleft
]{an}
\usepackage{graphicx}
\usepackage{times}
\begin{document}

\Pagespan{789}{}
\Yearpublication{2007}%
\Yearsubmission{2006}%
\Month{10}%
\Volume{999}%
\Issue{88}%

\title{Triple correlations in local helioseismology}

\author{F.P. Pijpers\inst{1}\fnmsep\thanks{
  \email{F.Pijpers@imperial.ac.uk}\newline}
}
\titlerunning{triple correlation helioseismology}
\authorrunning{F.P. Pijpers}
\institute{
SPAT, Dept. of Physics, Imperial College London, Blackett lab., 
Prince Consort Road, London SW7 2BW, England }

\received{15 Oct 2006}
\accepted{yes}
\publonline{later}


\abstract{%
  A central step in time-distance local helioseismology techniques is
  to obtain travel times of packets of wave signals between points or sets of
  points on the visible surface. Standard ways of determining group or phase
  travel times involve cross-correlating the signal between locations
  at the solar photosphere and determining the shift of the envelope 
  of this cross correlation function, or a zero crossing, using 
  a standard wavelet or a reference wave packet. Here a novel method
  is described which makes use of triple correlations,
  i.e. cross-correlating signal between three locations. By using an
  average triple correlation as reference, differential travel times
  can be extracted in a straightforward manner.}

\maketitle

\section{Introduction}
Since the first realisation of the usefulness of time-distance
helioseismology (\cite{Duv+93}) the technique has seen considerable 
application and has developed also in terms of the tools used to
interpret the relationship between the time delays measured for wave
packets travelling a specified horizontal distance and the
perturbations in temperature and velocity fields below the visible surface.
A recent review describing the current state of the subject is 
\cite{GizBir05}.

Time-distance helioseismic processing starts with the raw visible surface 
velocity data in the form of Dopplergrams and ends with performing the
inverse problem to obtain tomographic images of sub-surface
structures. An essential intermediate step is to carry out
cross-correlation of velocity time series between locations where the
relevant spectral line is formed, separated by one or more specified
horizontal distances. Normally the cross-correlation is carried out
between patches of pixels. e.g. centre-annulus or arc-to-arc
cross-correlation, in order to improve the signal-to-noise ratio.
Also, the signal is often pre-filtered, to let through only signals
travelling at horizontal phase speeds confined within a small band.
 
The cross-correlation function typically takes the form of a wavelet
centered at a certain delay time away from $0$, which is superposed on
noise. Therefore one can fit e.g. a Morlet wavelet to this in order to
extract the time of the maximum of the envelope, which is a proxy for
the group travel time, or the time of a selected zero-crossing which
is a proxy forn the phase travel time. These travel times are then
the input for the inverse problem. Recently, an improvement of this
technique has been proposed (\cite{GizBir04}) where instead of a
Morlet wavelet, a reference wavelet is used, for instance obtained
from quiet Sun data, or computed from a reference solar model.
The details of the fitting and time-delay extraction can be found 
in \cite{GizBir04}. This technique is more robust in the sense that it 
continues to function at lower signal-to-noise ratios. Nevertheless it
is still somewhat cumbersome and there is scope for alternative
estimators for travel times. In this paper one possible alternative is
discussed, which is based on triple correlations. In some sense the
technique is carried over from for instance speckle masking
interferometry where it has been applied succesfully for compensating
for atmospheric seeing to achieve aperture-limited high spatial
resolution imaging in a variety of settings (cf. \cite{Loh+83}, \cite{Bar+84}).

\section{Technique}
The triple correlation of three time series $f_1$ $f_2$ and $f_3$ is 
defined as~:
\begin{equation}
c(\tau_1, \tau_2) \equiv \int{\mathrm d}t\ f_1(t) f_2(t+\tau_1)
f_3(t+\tau_2)\ .
\end{equation}
This is most conveniently calculated in the Fourier domain where the
Fourier transform $C$ of the correlation function $c$ is related to
the Fourier transforms $F_1$, $F_2$ and $F_3$ of the three time series
by taking their product as follows~:
\begin{equation}
C(\omega_1, \omega_2) = F_1(\omega_1) F_2(\omega_2)
F_3^*(\omega_1+\omega_2)\ .
\end{equation}
Although it is possible to calculate these triple products directly 
from the Fourier transform of single pixel time series, in the first 
instance the pre-processing is retained from the standard two-point 
cross correlations, using averaging masks to improve the signal-to-noise 
ratio and a phase speed filter to isolate a wave packet. This is done 
in order to compare the results of the two methods, and also in order 
to assess the sensitivity of the results to noise propagation, by 
adjusting the parameters of the masks and filters.

\begin{figure}
\includegraphics[width=83mm]{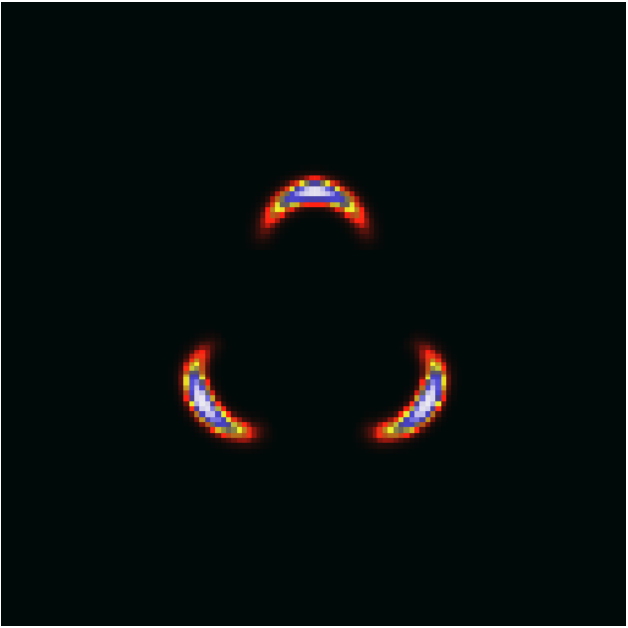}
\caption{The weights in the averaging masks for the triple
correlation shown as a linear color scale with red the smallest and white
the largest relative weight. Note that the masks are all slightly 
displaced outward from the origin to bring out the structure more clearly.}
\label{avgmasks}
\end{figure}

A GONG (Global Oscillations Network Group) data\-cube, tracked at the
solar rotation rate, and re-sampled onto a common spatial grid is used
for testing purposes. This cube is Fourier transformed in the two
spatial directions and in the time direction. The mask averaging is
done conveniently in the Fourier domain, because of efficiency. If one
needs to mask and average data using a mask with a fixed shape which
is shifted around to sample the entire domain available, this can be
expressed as a convolution. In the (spatial) Fourier domain the
convolution becomes a simple multiplication of the spatial Fourier
transform of the data and the Fourier transform of the mask. In this
way all necessary averages are obtained at once. Also, the volume of
data that is needed for subsequent steps is reduced drastically since
it is now significantly oversampled in the spatial directions so that a
lot of redundant data can be dropped from the subsequent steps in
the analysis.

Similarly the application of the phase speed filter to the data is done 
by Fourier transforming the phase speed filter and multiplying through 
with the Fourier transformed data-cube. The phase speed filter $H$ used is
Gaussian in shape~:
\begin{equation}
H({\bf k}, \omega) = \frac{2}{\Delta_v}\sqrt{\frac{\ln 2}{\pi}} \exp\left[ - 
\frac{4\ln 2}{\Delta_v^2}\left(\frac{\omega}{\vert k \vert} - v_{\rm cen} 
\right)^2\right]\ .
\end{equation}
With this definition $\Delta_v$ corresponds to the full-width at half 
maximum (FWHM).
The masks chosen for the three point correlation function are
arcs. The centres of the three arcs are placed on an equilateral
triangle at a distance corresponding to $29\ {\mathrm{Mm}}$. For the
purposes of these feasibility tests only a single distance is
used. For full tomographic inversions, a range of distances would be
used, and the properties of the phase speed filter and mask would be
adjusted accordingly. The weighting is not uniform over the arcs~: the
profile of the arc in the radial direction as well as in the azimuth
angle is Gaussian, with a FWHM of $10\ {\mathrm{Mm}}$ and $90^o$
respectively. The masks are illustrated in Fig. \ref{avgmasks} which
shows as a color scale the averaging weight for the three arcs. Each
arc is normalised to have a unit sum over the pixel weights. The
parameters of the phase speed filter are set to~:
\begin{eqnarray}
v_{\rm cen} &&= 44\ {\mathrm{km/s}}\nonumber\\
\Delta_v &&= 17.6\ {\mathrm{km/s}}\ .
\end{eqnarray}

\begin{figure}
\includegraphics[width=60mm]{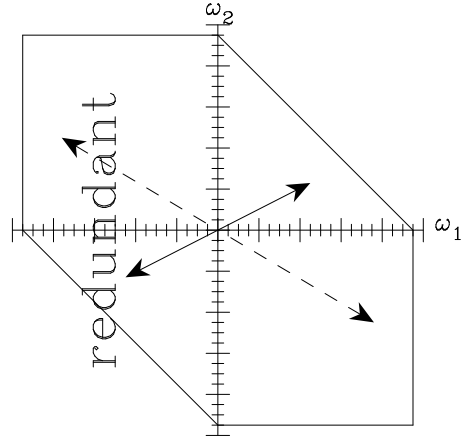}
\caption{In the Fourier domain a triple correlation fills a hexagonal
area, bounded by the Nyquist frequency in $\omega_1$ $\omega_2$ and 
$\omega_1+\omega_2$, shown in the figure. Since the time series are
real there is redundancy and only two quadrants need be manipulated.
The arrows indicate which points are complex conjugate pairs.}
\label{diagtripcor}
\end{figure}

Before discussing the results of the triple correlations it is useful
to consider what one would expect. The time series are sampled at a
finite rate, which implies that there is no information in the Fourier
domain for frequencies that are in absolute value above the Nyquist 
frequency. In the domain of the two Fourier frequencies $\omega_1,
\omega_2$ (see Fig. \ref{diagtripcor}) one can therefore restrict the
analysis to a square region centered on the origin. Two triangles are
further `cut away' because these fall outside the Nyquist range for 
$\omega_1+\omega_2$. This hexagonal area contains the complex valued
Fourier transform of the triple correlation function. Since the 
original time series, and therefore also their triple correlation, is
real valued, points in this diagram that are mirror images with
respect to the origin (see the arrows in Fig. \ref{diagtripcor}) are
complex conjugates. 
\begin{equation}
C(-\omega_1, -\omega_2) \equiv C^*(\omega_1, \omega_2)
\end{equation}
There is therefore no need to retain all 4
quadrants~: two suffice and here the choice is made to retain quadrants 1
and 4. In speckle interferometry there are further symmetries in the
quantities that are being correlated, with the consequence that the 
domain necessary to fully specify the information can be cut down
further. That is not the case here~; the time series are not invariant
under time reversal for instance, and therefore no further reduction
is possible.

In the time domain, two-point correlation functions tend to have a
wavelet shape superposed on noise. In the Fourier domain, two-point 
correlation functions are normally also structured with multiple
peaks, and three point correlation functions therefore are as well. 
Most commonly in time-distance helioseismology, the shape of the
wavelet is not used. The only parameter that is extracted is a time
delay, either from a zero crossing or from the location of the maximum
of an envelope. It is consistent with this to assume that to lowest
order the shape of the triple correlation i.e. its complex modulus, does not
change over a field of triple correlations. To this order, the only
quantity of interest is the relative displacement of this wavelet to
larger or smaller delays, which in the Fourier domain corresponds to a
shift in the complex phase. In other words the triple correlation
wavelet $w(\tau_1, \tau_2)$ in the time domain is presumed to be 
related to an (ensemble) average $\langle w \rangle(\tau_1, \tau_2)$ as~:
\begin{eqnarray}
w(\tau_1, \tau_2) = \int{\rm d} \tau^\prime_1\int{\rm d}\tau^\prime_2 
&&\langle w \rangle(\tau^\prime_1, \tau^\prime_2) \times\nonumber\\
&&\,\delta(\tau^\prime_1-\tau_1)\delta(\tau^\prime_2-\tau_2)\ ,
\end{eqnarray}
in which the $\delta$ are Dirac delta functions. This describes a
convolution and therefore can be expressed as a simple multiplication
in the Fourier domain. This suggests an approach such as Wiener
filtering, in which the Fourier transform of the triple correlation
$W$ is divided by the Fourier transform of the ensemble averaged triple
correlation $\langle W \rangle$. Then what one retains is the Fourier 
transforms of the two $\delta$-functions i.e. a function of the form
$e^{i\phi}$ in which the complex phase $\phi$ is~:
\begin{equation}
\phi = \omega_1\Delta_1 + \omega_2 \Delta_2\ .
\end{equation}
The $\Delta_1$ and $\Delta_2$ are differences in travel time between
and individual triple correlation and the average, as represented by 
the average triple correlation $\langle w \rangle$. 

\begin{figure}
\includegraphics[width=83mm]{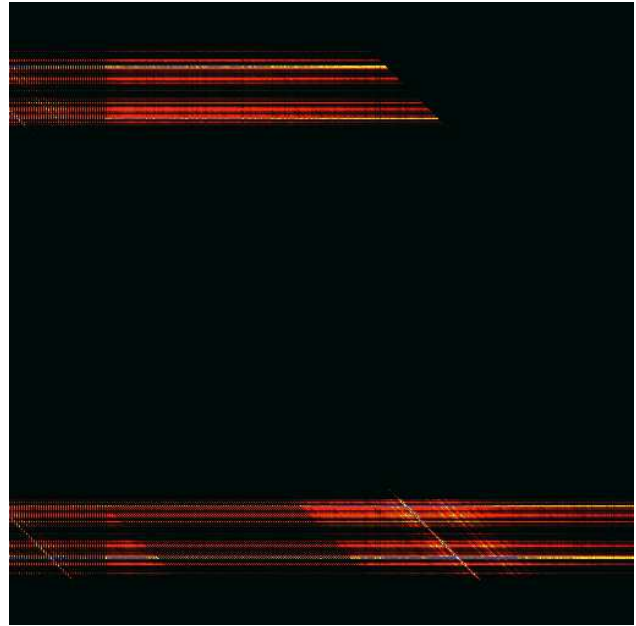}
\caption{The modulus of the complex valued Fourier transform of the
triple correlation in quadrants 1 and 4, when averaged over the entire
$8\times 8$ field of triple correlations. The color scale is logarithmic 
covering 20 powers of 10~; black and red representing smallest and
blue and white largest values.}
\label{avgmod}
\end{figure}

\begin{figure}
\includegraphics[width=83mm]{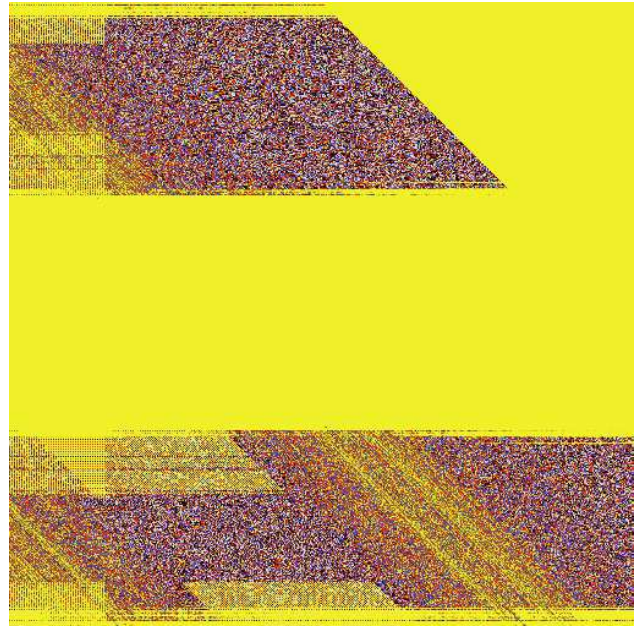}
\caption{The complex phase of the Fourier transform of the
triple correlation in quadrants 1 and 4, when averaged over the entire
$8\times 8$ field of triple correlations. The color scale is linear : yellow
is zero phase, red and black is negative phase, blue and white is
positive phase.}
\label{avgpha}
\end{figure}

In Figs. \ref{avgmod} and \ref{avgpha} the complex modulus and phase
are shown of the average triple correlation in the Fourier domain 
$\langle W \rangle$ obtained by averaging the Fourier amplitudes and
phases over all of the $8\times 8$ triple correlations over the field
covered by the data-cube. The phase is set to $0$ where the modulus is below a
certain threshold. By construction the distances between the three
points being correlated is the same and therefore one would expect the
maximum correlation to occur for $\tau_1 = \tau_2$ in the average
correlation $\langle w \rangle$. In the Fourier domain this
corresponds to structures aligned along $\omega_1+\omega_2 =
\mathrm{const.}$. For the same reason the individual cross
correlations and the ratios $W/\langle W \rangle$ should show the same
structure. This is particularly clear in Fig. \ref{avgpha}. In Fig. 
\ref{avgmod} the modulus is high only in very localised regions so
that such a structure is lost.

In calculating the ratio of triple correlations there is evidently a
problem in large regions of the domain where the average 
$\langle W \rangle =0$.
This is not an uncommon problem in inversions, and one expression of
the need for regularisation in inverse problems. In this case the most
straightforward regularisation is similar to what is done in singular
value decomposition. Rather than using $W/\langle W\rangle$ one uses~:
\begin{equation}
\Phi (\omega_1, \omega_2) = \frac{W(\omega_1,
\omega_2)}{\langle W(\omega_1, \omega_2)\rangle + \epsilon}
\end{equation}
where $\epsilon$ is a suitably small number which needs to be adjusted
to the problem at hand. Here $\epsilon = 10^{-13}$, which is small
compared to the maximum value in the average triple correlation 
$\langle W\rangle_{\mathrm{max}} \sim 10^{16}$.

\begin{figure}
\includegraphics[width=83mm]{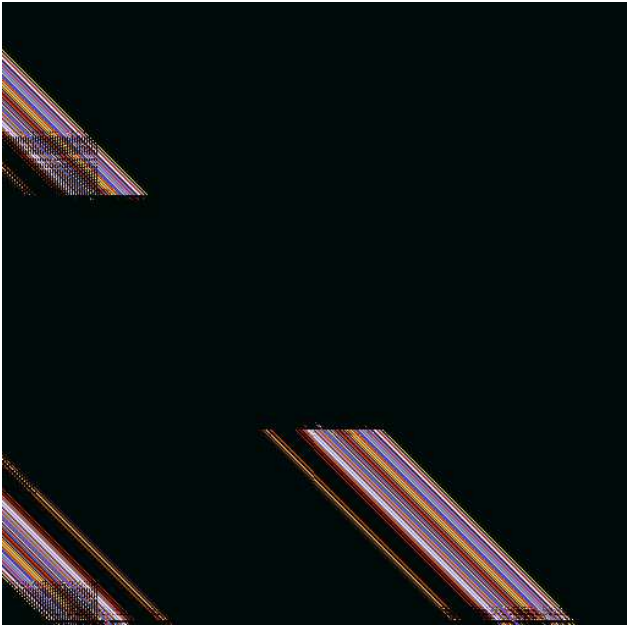}
\caption{The modulus of the complex valued Fourier transform of the
triple correlation ratio $\Phi$ in quadrants 1 and 4, at a location
near the centre of the field. The color scale is logarithmic 
covering 5 powers of 10 with color coding as in Fig. 3.}
\label{locmod}
\end{figure}

\begin{figure}
\includegraphics[width=83mm]{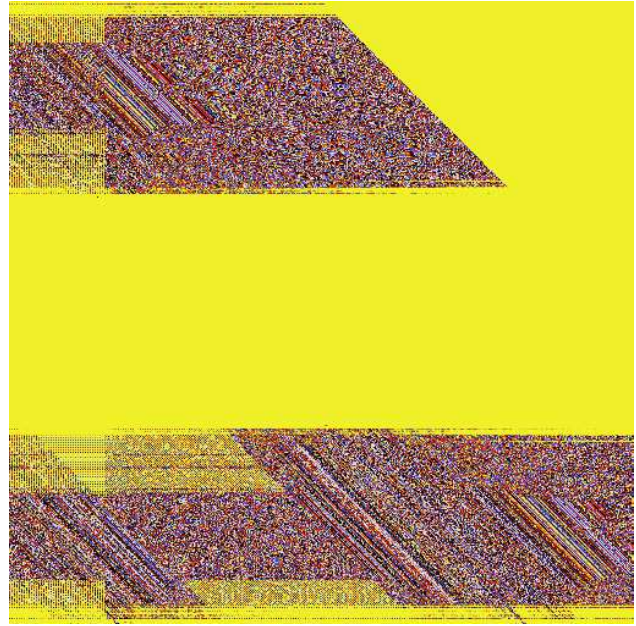}
\caption{The complex phase of the Fourier transform of the
triple correlation ratio $\Phi$ in quadrants 1 and 4, at a location
near the centre of the field. The color scale is linear : yellow
is zero phase, red and black is negative phase, blue and white is
positive phase.}
\label{locpha}
\end{figure}

\begin{figure}
\includegraphics[width=83mm]{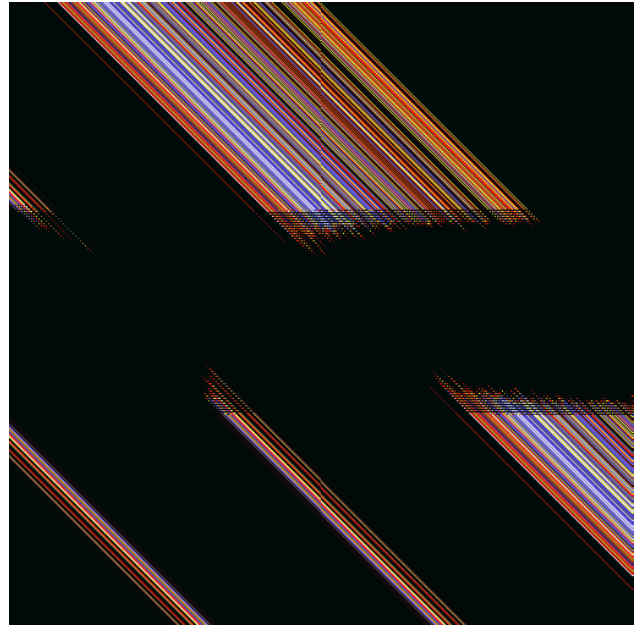}
\caption{The modulus of the complex valued Fourier transform of the
triple correlation ratio $\Phi$ in quadrants 1 and 4, at the same 
location as Fig. 5 but without any filtering. The color scale is 
logarithmic covering 5 powers of 10 with color coding as in Fig. 3.}
\label{locmodnof}
\end{figure}

\begin{figure}
\includegraphics[width=83mm]{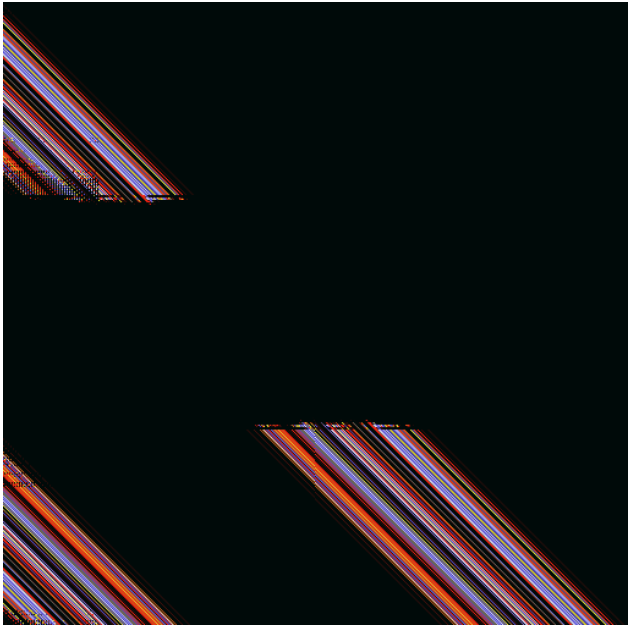}
\caption{The modulus of the complex valued Fourier transform of the
triple correlation ratio $\Phi$ in quadrants 1 and 4, at the same 
location as Fig. 5 but with a more localised mask. The color scale is 
logarithmic covering 5 powers of 10 with color coding as in Fig. 3.}
\label{locmodsmm}
\end{figure}

If the shape of the wavelet were indeed constant over the field, and
only displaced to larger or smaller relative time delays, the maps
of the complex modulus of $\Phi$ would be featureless~: equal to $1$ 
everywhere. The complex phase of $\Phi$ would show tructures aligned
along $\omega_1+\omega_2 = \mathrm{const.}$ as mentioned above.
However, the group and phase travel time of waves both change over the
field, but not necessarily by the same value, due to the fact that the
dispersion relation is complex. Only if the sub-photospheric region
of the Sun were to satisfy~:
\begin{equation}
\frac{\vert\bf  k\vert}{\omega}\frac{\partial\omega}{\partial\vert\bf
k\vert} = \mathrm{const.}
\end{equation}
would $\Phi$ show this behaviour. Since this is not the case, the
modulus of $\Phi$ also shows structure, as does its phase. 

In Figs. \ref{locmod} and \ref{locpha} the triple correlation ratio
in the Fourier domain $\Phi$ is shown for a single location near the
centre of the field. Fig. \ref{locmod} shows the logarithm of the
modulus and Fig. \ref{locpha} shows the phase. There are two clear
ridges present as expected, both in the modulus and in the phase
diagrams, from which a time-delay can be extracted. It is clear that
along rdiges of $\omega_1+\omega_2=\mathrm{const.}$ there is little
variation other than what is created because of the regularisation
process. One can therefore integrate along the ridges and reduce the
time-delay extraction to determining a single mean value $\tau_1 =
\tau_2 = \tau_{\mathrm{m}}$ from the remaining 1-D Fourier transform.

\section{Discussion \& Conclusions}
The triple correlation technique to extract relative time delays is 
demonstrated to function on a data-set obtained from GONG. The tests
discussed in this paper are done on real data for an active region, 
with standard filtering and averaging applied and therefore it is
evident that noise does not pose a significant problem in extracting
time delays. However, further work is necessary to establish the
relationship between the relative  time delays recovered in this way,
and the perturbations in sub-photospheric layers. As has been pointed
out in \cite{GizBir02} and in \cite{JenPij03} it is possible to
express this relationship in terms of a linear inverse problem with
known kernels, as long as perturbations from a known equilibrium are
small. However, the precise form that these kernels take does depend
on the processing in terms of averaging and filtering. Although the
differences with existing kernels are unlikely to be substantial,
appropriate kernels for the triple correlation processing are still to
be derived. 

A point to note is that the time delays extracted are measured
relative to the mean solar sub-photosphere over the time period
covered by the data. This mean is not necessarily identical to a
standard solar model. In order to be able to interpret the time delays
with the appropriate kernels calculated from a standard model one
should also extract the differential time delay of the average with
respect to the standard solar model. This can be done by calculating
the triple correlation $W_{\mathrm{mod}}$ as it would be for the model
and extracting the delay from the (regularised) ratio $\langle W
\rangle/(W_{\mathrm{mod}} + \epsilon)$. The reason to proceed in this
way is that the average triple correlation $\langle W \rangle$ is
a horizontal average for the region, for which a comparison with a
(horizontally invariant) standard solar model is more straightforward
to interpret. The horizontal perturbations measured by individual
triple correlations are more likely to be linear when compared with a
horizontal average, than they are when compared with a standard solar model.

In order to illustrate the sensitivity to the filtering of the triple
correlations, the same data-cube was processed but without using any
filtering at all. The equivalent of Fig. \ref{locmod} is shown in
Fig. \ref{locmodnof}. It is clear that qualitatively similar structure
is present in both images so that even for unfiltered data it would be
possible to extract a relative time delay. However in the unfiltered
processing the relative delays would be quite different from those 
recovered from filtered data. This is due to the fact that
without the filtering there is a mixture of wave modes with very
different depths of penetration into the sub-surface layers. The
relative delays with respect to the mean appear to be much larger
here, which suggests that these data are dominated by modes that
remain in shallow layers, such as f modes. The influence of noise,
either measurement noise or intrinsic solar noise, appears to be
minimal. 

A separate test is to retain the filtering, but to reduce the extent
of the averaging mask. In Fig. \ref{locmodsmm} both the radial extent
and the azimuthal extent (FWHM) of the arcs is reduced by a factor of
$\sqrt{2}$. This figure shows essentially the same structure as
Fig. \ref{locmod} although perhaps more of the expected periodic
structure is present in the form of a third ridge, which implies that
the time delay is somewhat better constrained for the smaller
mask. Further optimisation of the combination of filtering and spatial
averaging is in progress. 

The computational burden of processing this data cube of 512 slices of
128x128 pixels is $12\ \mathrm{min}$ on an 8 processor machine. This
is very similar to the standard two point correlation processing, with
a more direct and robust access to the time delays. The preliminary
tests with filtering and averaging indicate that perhaps less
averaging is necessary compared to standard two-point correlation
functions, which would be beneficial for subsequent tomography.

\acknowledgements
This work utilizes data obtained by the Global Oscillation Network
Group (GONG) program, managed by the National Solar Observatory, which
is operated by AURA, Inc. under a cooperative agreement with the
National Science Foundation. The data were acquired by instruments
operated by the Big Bear Solar Observatory, High Altitude Observatory,
Learmonth Solar Observatory, Udaipur Solar Observatory, Instituto de
Astrof\'{\i}sica de Canarias, and Cerro Tololo Interamerican
Observatory.

\newpage

\end{document}